\documentclass{ws-p8-50x6-00}
\begin{document}

\title{Magnetic Fluxes, NS-NS $B$ Field and Shifts in
Four-Dimensional Orientifolds\thanks{{\it  ROM2F/2003/26}}}

\author{Gianfranco Pradisi}

\address{ Dipartimento di Fisica, Universit\`a di Roma
``Tor Vergata''\\INFN,
Sezione di Roma ``Tor Vergata''\\ \sl
Via della Ricerca Scientifica 1, I-00133 Roma,
Italy\\
E-mail: gianfranco.pradisi@roma2.infn.it}


\maketitle

\abstracts{We discuss the interplay between freely acting orbifold
actions, discrete deformations and internal uniform magnetic
fields in four-dimensional orientifold models.}

\section{Introduction}

In the framework of string/M-theory attempts\cite{rev} to
single-out models with a particle content as close as possible to
(some extension of) the Standard Model, a prominent role is played
by type I vacua\cite{tipo1,bs} with uniform internal magnetic
fields\cite{mag6,ted,magn}, or, in a T-dual language, with
$D$-branes intersecting at angles\cite{rev2}. In this talk, we
review the basic ingredients entering these constructions,
stressing their most appealing features, namely chirality, the
natural emergence of replicas of matter-field families and the
surprising power of combinations of shifts in freely acting
orbifolds and the NS-NS $B_{ab}$, itself equivalent to an
asymmetric shift-orbifold projection.  In particular, after a
general discussion, we show an example that displays all these
properties, extracted from ref. (\cite{magn}), where an exhaustive
discussion of four dimensional $Z_2 \times Z_2$ shift-orientifolds
can be found.
\section{Shift-Orientifolds}

The simplest example of a shift orbifold can be obtained starting
from a one-dimensional boson $X$ compactified on a circle of
radius $R$ and identifying its values under the (freely acting)
antipodal transformation, namely $X \rightarrow X + \pi R$.
Twisted sectors depend on $R$, and the orbifold procedure, not
surprisingly, yields simply a one-dimensional boson compactified
on a circle of radius $R/2$.  However, the combination of
freely-acting actions with internal (discrete) symmetries realizes
in string theory the Scherk-Schwarz mechanism\cite{ss}, via the
assignment of different periodicity conditions to bosons and
fermions along the internal directions, giving rise to the
(spontaneous) breaking of supersymmetry.\cite{closs} In the
framework of type II superstrings, a T-duality can turn a shift of
internal momenta into a shift of internal windings, that therefore
exhibit analogous features. In orientifolds, however, things are
quite different.  Indeed, momentum shifts are conventional
Scherk-Schwarz deformations, while winding shifts can be related
via T-duality to momentum shifts along the $11$-th coordinate of
the underlying M-theory. Alternatively, the ``massless'' spectrum
depends crucially on the relative orientations of the momentum
shifts and the $D$-branes present. Thus, a shift longitudinal to
the $D$-branes in a certain sector gives rise to a conventional
Scherk-Schwarz deformation, with supersymmetry already broken at
tree level in string perturbation theory.  On the contrary, with a
momentum shift orthogonal to the $D$-branes in a certain sector,
the supersymmetry of the massless modes (and sometimes also of the
massive ones) is not affected by the orbifold action. This
phenomenon, termed ``brane supersymmetry'' in the
literature\cite{brasusy}, has a very neat geometric
interpretation: it can be ascribed to configurations with {\it
multiplets} of $D$-branes.  Indeed, an orthogonal momentum shift
is compatible with a brane configuration if, and only if, for each
given $D$-brane at a certain ``position'' in spacetime, there
exist image $D$-branes at each image ``position'' obtained from
the initial one. Multiplets of branes are a generic feature of
shift-orientifolds.
\begin{table}[t]
\caption{$Z_2 \times Z_2$ $4$-d shift-orientifolds, $D$-branes and
``brane supersymmetry''.}
\begin{center}
\footnotesize
\begin{tabular}{|c|c|c|c|}\hline
{\rm models}&{$D9$ susy}&{$D5_{1}$ susy}&{$D5_{2}$ susy}
\\ \hline
{$p_3$}&{N=1}&{N=2} &{N=2}\\
{$w_2p_3$}&{N=2}&{N=2} &{N=4}\\
{$w_1w_2p_3$}&{N=4}&{N=4}&{N=4} \\  \hline
{$p_2 p_3$}&{N=1}&{N=2}&{--} \\
{$w_1p_2$}&{N=2}&{N=4}&{--} \\
{$w_1p_2p_3$}&{N=2}&{N=4}&{--} \\
{$w_1p_2w_3$}&{N=4}&{N=4}&{--} \\ \hline
{$p_1 p_2 p_3$}&{N=1}&{--}&{--} \\
{$p_1w_2w_3$}&{N=2}&{--} &{--}\\
{$w_1w_2w_3$}&{N=4}&{--}&{--}
\\ \hline
\end{tabular}
\label{oldmodels}
\end{center}
\end{table}
A very interesting class of these models in four dimensions,
extensively studied in refs. (\cite{adds}) and (\cite{aadds}), corresponds to
orientifolds of freely acting orbifolds obtained combining $Z_2
\times Z_2$ inversions and shifts.  More precisely, the six torus
is taken as a product of three two-tori $T^2 \times T^2 \times
T^2$, while the orbifold action is a combination of the $Z_2
\times Z_2$ group containing the identity and the three inversions
of pairs of the three complex torus coordinates, combined with
shifts along their real parts. Up to T-duality and trivial
redefinitions, the types of inequivalent models are collected in
table~\ref{oldmodels}, together with their degree of ``brane
supersymmetry'', that signals the brane-multiplet structure of the
corresponding orientifolds. Their open spectra are very rich but
unfortunately are not chiral, and can be found in refs.
(\cite{adds}) and (\cite{aadds}).
\section{$B_{ab}$ and Shifts}

The presence of a NS-NS two form field $B_{ab}$ in type I vacua
has far reaching consequences that induce several interesting
effects. The orientifold projection is actually compatible only
with {\it quantized} values of the background field, a feature
that can be easily deduced analyzing the type IIB superstring
compactified on a $d$-dimensional torus\cite{bps}. The generalized
left and right momenta entering the Narain lattice can be written
in the form 
\bea 
p_{{\rm L},a} = m_a + \frac{1}{\alpha'} ( g_{ab}
- B_{ab} ) \, n^b \, , \nonumber 
\\
p_{{\rm R},a} = m_a - \frac{1}{\alpha'} ( g_{ab} + B_{ab} ) \, n^b
\, , 
\label{momwind} 
\eea 
where $m_a$ and $n^a$ are
the (integer) momentum and winding quantum numbers, respectively.
The Narain lattice admits a sensible action of the world-sheet
parity operator $\Omega$ only if the condition 
\be
\frac{2}{\alpha'} \ B_{ab} \ \in \ {\b{Z}} 
\ee 
holds\footnote{If the
orientifold projection is a combination of $\Omega$ and an
involution, the discretization affects other moduli, for instance
some off-diagonal components of the torus metric\cite{anblu}.}:
$B_{ab}$ must belong to the integer cohomology of the
$d$-dimensional torus and the independent values of its components
are $0$ and $1/2$ in suitable inverse $\alpha'$ units.
Equivalently, the quantized values of $B_{ab}$ can be ascribed to
a toroidal compactification {\it without} vector structure. Indeed, the
deep connection between the bulk (closed) sector and the boundary
(open) sector in orientifolds reflects itself in the
identification\cite{senset} of the quantized $B_{ab}$, a
theta-like angle of the closed-string sector, with the mod-two
cohomology class $\tilde{w}_2$ that, in analogy with the
Stiefel-Whitney class $w_2$ for spin-bundles, measures the
obstruction to defining a vector structure on the vacuum gauge
bundle\cite{seiaut}. Gauge fields are in the brane sectors of
orientifolds, but their numbers and representations depend on the
presence of non-trivial $B_{ab}$ fluxes. The most striking effect
of a non-vanishing $B_{ab}$-field on orientifolds is the
reduction of the rank of the Chan-Paton gauge group by a factor of
$2^{r/2}$, where $r$ is the rank of $B_{ab}$.  In the original
derivation\cite{bps}, the rank reduction was deduced
algebraically considering the one loop partition function at
rational points in the moduli space, where the quantized $B_{ab}$
was implicit in the Narain lattice choice\cite{bs}, and at generic
radii, imposing the correct normalization of the closed states
along the tube in the annulus and M\"obius amplitudes.
The topological properties of the toroidal compactification
without vector structure allow an alternative interpretation in
terms of non-commuting Wilson lines. These feel the $\tilde{w}_2$,
leading to a non-commutativity of the Chan-Paton gauge matrices,
that reduces by the mentioned factor $2^{r/2}$ the rank of the
Chan-Paton group\cite{senset,kaku2,maxtor,wittor,kst}. $r$ is also
the parameter that discriminates between disconnected components
of the moduli space of type I compactifications with $16$
supercharges\cite{maxtor}.

Geometric interpretations can be given in terms of the extended objects
present in the target space of type I vacua, namely $D$-branes and $O$-planes.
Without loss of generality, we can limit ourselves to the case of a
two-torus equipped with space-filling $O_{2+}$-planes and stacks of
N $D_2$ branes, that correspond to configurations of standard
toroidal compactifications {\it with} vector structure and a standard
$\Omega$ orientifold projection.  More complicated cases exist, but the
qualitative features remain the same.
\begin{figure}[t]
\begin{center}
\epsfxsize=12cm %
\epsfbox{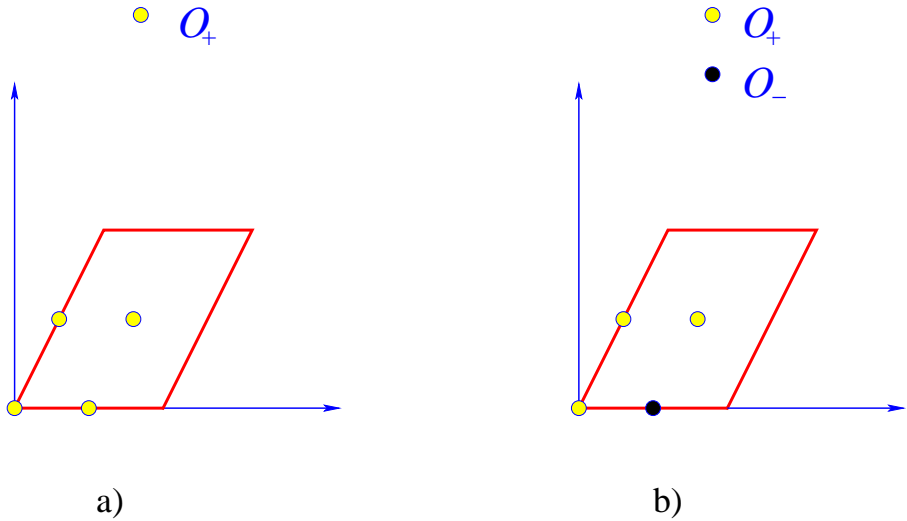}
\caption{$O_0$-planes configuration on a two-torus: a) $B=0$. b) $B \ne 0$
\label{opiani}}
\end{center}
\end{figure}
It was observed in ref. (\cite{wittor}) that, performing a pair of
T-dualities along the two directions of the torus, one ends up
with a configuration exhibiting a quartet of $O_{0+}$-planes
located at the fixed points of a $Z_2$ orbifold of the two-torus
itself, so that, say, N $D$-branes are required to neutralize the
overall R-R charge of the configuration.  On the other hand, if
the two-torus is without vector structure, two different types of
$O$-planes are present, the conventional $O_+$'s with tension and
charge opposite to those of the $D$-branes, and ``exotic''
$O_-$-planes with reverted tension and charge. As a result, the
net charge of the configuration is reduced by a factor of two,
because after two T-dualities one is faced with three $O_+$-planes
and one $O_-$-plane (see fig.~\ref{opiani}), in a configuration
that requires a net number of N/2 $D$-branes (as compared to N,
the number previously required) to neutralize the total R-R
charge. All this reflects precisely the features of partition
functions in ref. (\cite{bps}).

\begin{figure}[t]
\begin{center}
\epsfxsize=12cm %
\epsfbox{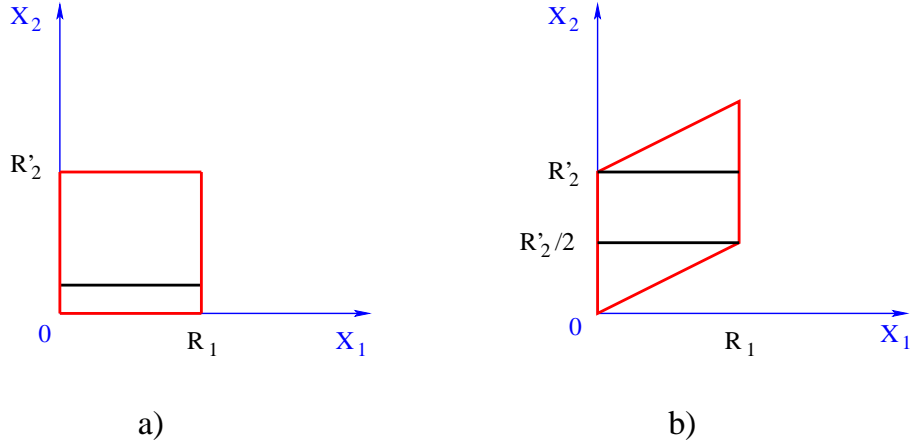}
\caption{$D_1$-brane configuration of the $\Omega \ R$ orientifold:
a) $B=0$. b) $B \ne 0$
\label{tilt}}
\end{center}
\end{figure}

An interesting alternative geometrical interpretation, useful in
the context of intersecting $D$-brane models, can also be given as
follows.  Let us start, for clarity, with a rectangular torus of
sides $R_1$ and $R_2$, and let us perform a T-duality along the
vertical direction in such a way that the complex and K\"ahler
structures get interchanged.  The result depends crucially on the
vector structure of the torus.  Indeed, as shown in
fig.~\ref{tilt}, the $O_2$'s and the $D_2$'s becomes $O_1$'s and
$D_1$'s after the T-duality, but the shape of the final torus is
very different in the two cases: if $B_{ab}=0$, one is left with a
new rectangular torus of sides $R_1$ and $R'_2 = \alpha'/R_2$. On
the contrary, if $B_{ab} \ne 0$, one ends up with a tilted torus
admitting vector structure, i.e. with  $B'_{ab}=0$.
However, in the latter case, each of the resulting $D_1$-branes
has twice the minimal length occurring in the $B_{ab}=0$ case. The
independent branes are thus half of their na\"ive
 number, and this provides the expected rank
reduction, consistently with the doubling of their elementary
charge.

\begin{figure}[t]
\begin{center}
\epsfxsize=12cm %
\epsfbox{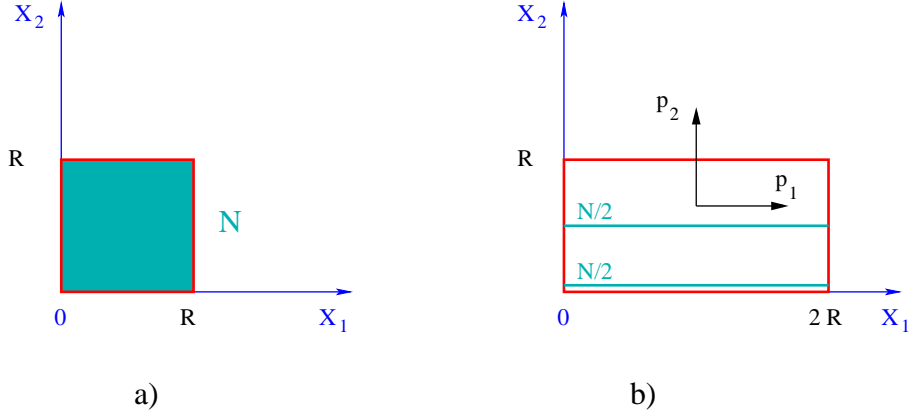}
\caption{$B$-field vs shift-orientifold: a) space filling
$D_2$-branes with $B=0$. b) Doublets of $D_1$-branes in the shift-orientifold.
\label{bshift}}
\end{center}
\end{figure}

In the spirit of section 2, another nice interpretation of the
rank reduction for the Chan-Paton group rests on the observation
that a quantized $B_{ab}$ is equivalent to a toroidal
shift-orbifold compactification\cite{magn,kaku}. Let us consider
for simplicity a two-torus that is the product of two circles of
radius R, and let us choose $B_{ab}$ in eq. (\ref{momwind}) so
that the two-dimensional blocks are parametrized as 
\bea 
B \ = \
\frac{\alpha'}{2} \left(
\begin{array}{rrrr}
0&1\\
-1&0\end{array} \right) \quad .
\label{bdiscrete}
\eea
The non-trivial background makes
the sum over momenta dependent on the parity of the integer windings,
in such a way that  the one-loop torus partition function can be decomposed
in the form
\bea
&\Lambda&\!\!(B) \ = \ \Lambda(m_1, m_2, 2n_1, 2n_2) +
\Lambda(m_1+1/2, m_2, 2n_1, 2n_2+1) + \label{torob} \\
&\Lambda&\!\!(m_1, m_2+1/2, 2n_1+1, 2n_2) + \Lambda(m_1+1/2,
m_2+1/2, 2n_1+1, 2n_2+1) \, , \nonumber 
\eea 
where $\Lambda(m_1,
m_2, n_1, n_2)$ denotes the two-dimensional lattice sum over
momenta $m_a / R$ and windings $n^a R$. This expression is indeed
the partition function of an asymmetric $p_1w_2$ shift-orbifold of
a two-torus with $B_{ab}=0$, provided the radius along the first
direction is doubled $(R \rightarrow 2 R)$.  As a result, the
toroidal compactification on the two-dimensional torus without
vector structure is equivalent to a $p_1w_2$ shift-orbifold
compactification, but on a different torus, characterized by a
fundamental cell of double length along the first direction.  In
order to exhibit the expected rank reduction, it is useful to
study the $\Omega$-orientifold after performing a T-duality along
the vertical direction, that renders the $p_1w_2$ shift-orbifold
action equivalent to a $p_1 p_2$ shift-orientifold where the
projection is combined with an inversion. The unoriented
truncation forces $m_2$ and $n_1$ to vanish in the Klein-bottle
amplitude, while $m_1$ and $n_2$ vanish along the tube, as
demanded by the properties of the corresponding boundary states\cite{croco}.
As a result, the Klein-bottle amplitude takes the form 
\be 
{\cal K} \ = \ \frac{1}{2} \ P_1 \ W_2 \ , 
\label{kleinp1p2} 
\ee 
where
$P_i$ and $W_i$ are the usual one-dimensional momentum and winding
sums\cite{tipo1,bps}, obtained as unoriented slices of the Narain
lattice, while the transverse annulus amplitude 
\bea 
{\tilde{\cal
A}} \ = \ \frac{ 2^{-3}}{2} \ N^2 \ \frac{v_1}{v_2} W_1^e \ P_2^e
\ , 
\label{aneltrap1p2} 
\eea 
contains sums over even momenta and
windings, weighted with the normalization needed for a proper
particle interpretation of the gauge vectors.  
The direct-channel
annulus amplitude, corresponding to the open-string one-loop
channel, can be written 
\be 
{\cal A} \ = \ \frac{1}{2} \ N^2 \
\big( \, P_1 + P_1^{1/2} \, \big) \ \big( \, W_2 + W_2^{1/2} \,
\big) \ , 
\label{anelp1p2} 
\ee 
where $P_i$ and $W_i$ are again the
usual one-dimensional momentum and winding lattice sums\cite{tipo1,bps}, 
respectively, while $P_i^{1/2}$ and $W_i^{1/2}$
are the corresponding shifted ones. 
The consistent M\"obius
amplitude, responsible for the unoriented projection, is then 
\be
{\cal M} \ = \ - \frac{1}{2} \ N \ \big( \, \hat{P}_1 \, \hat{W}_2
\, + \, \hat{P}_1^{1/2} \, \hat{W}_2^{1/2} \, \big) \ ,
\label{moebp1p2} 
\ee 
as can be checked via a $P$-transformation of
the transverse amplitude 
\be 
\tilde{\cal M} \ = \ - \, N \
\frac{v_1}{v_2} \ \left( \hat{W}_1^e \ \hat{P}_2^e \ + \ (-1)^{n_1
+ m_2} \ \hat{W}_1^e \ \hat{P}_2^e \ \right) \quad ,
\label{tramoep1p2} 
\ee 
demanded by eq. (\ref{aneltrap1p2}) and the
transverse Klein-bottle amplitude 
\be 
\tilde{\cal K} \ = \
\frac{2^{5}}{2} \ \frac{v_1}{v_2} \ W_1^e \ P_2^e \ .
\label{trakleinp1p2} 
\ee 
The presence of the shifted sums in eqs. 
(\ref{anelp1p2}) and (\ref{moebp1p2}) signals quite neatly that
one of the shifts, $p_2$ in this case, is orthogonal to the
$D_1$-branes, parallel to the $O_1$-planes and thus horizontal. 
As already observed is Section $2$, the orthogonal shift is
responsible for the doublet structure of the $D_1$-branes, forcing
the split into two stacks of $N/2$ brane-image doublets. The
geometrical situation is illustrated in fig.~\ref{bshift}. 
The consequent rank reduction of the Chan-Paton gauge group can be
easily derived from the analysis of the tadpole cancellation
conditions, or simply by inspection of eqs. (\ref{aneltrap1p2}),
(\ref{tramoep1p2}) and (\ref{trakleinp1p2}). For instance, the
previous construction, if applied to an eight-dimensional toroidal
compactification of the type IIB superstring, would result in type
I models with an $SO(16)$ Chan-Paton group and the expected rank
reduction by a factor of two.

Because of the direct connection between quantized $B_{ab}$ and
shift-orbifolds, it can happen that some combination of the two
makes $B_{ab}$ uneffective \cite{magn}.  In addition, a quantized
$B_{ab}$ generically introduces an open-string analog of discrete
torsion, whose values discriminate between Orthogonal and
Symplectic gauge groups. However, the two choices can be
continuously connected, passing through unitary groups via suitable
open-string Wilson lines\cite{bps}.  In
orbifolds\cite{kst,carlobab}, the quantized $B_{ab}$ also affects
the $\Omega$ eigenvalues of the twist fields, introducing again
exotic $O_-$-planes responsible for the different unoriented
truncations, according to the constraints dictated by the
underlying Conformal Field Theory\cite{croco}. At the same time,
additional matter multiplets emerge in the gauge sector of the
models in order to balance the loss of R-R charge of the
$O_-$-planes.
\section{Magnetized Branes and Chiral Models}

Chirality in four-dimensional string or M-theory inspired vacua is
not an easy property to obtain. The reason is, roughly speaking,
that the long wavelength modes come unavoidably from
higher-dimensional spinors, that generically give rise 
to vector-like configurations. Chiral models,
however, can be built using projections that retain a net number
of chiral spinors. The first example of chiral asymmetry in type I
vacua was exhibited in a class of orientifolds of the 
$Z$-orbifold\cite{chiral}.  In those examples, the $D9$-branes sit at singular
points of the internal manifold. However, chirality can also be
obtained via other mechanisms.  Widely studied in the past few
years are backgrounds equipped with uniform magnetic fields along
some internal directions of the branes\cite{ft,bachas,penta} or, in a
T-dual language\cite{intbranes}, with $D$-branes intersecting at
non-trivial angles\cite{rev2}. Magnetic fields couple solely to
the ends of open strings, interpolating between Dirichlet and
Neumann boundary conditions\cite{ft}. In compact manifolds, the
Dirac quantization condition obeyed by the magnetic charges is
equivalent to a rational (non-dense) wrapping of the magnetized
$D$-branes around corresponding $p$-cycles, counted exactly by the
degeneracy of the Landau levels. As a result, a fine-tuning of the
angles or of the background field-strengths to (anti-)self-dual
configurations allows to project away tachyonic modes and to
restore supersymmetry, generically broken by the couplings of
different spins to the magnetic field. Moreover, due to the
non-vanishing topological charge, magnetized branes transmute to
uncharged branes of diverse dimensionalities, {\it i.e.} they
couple to R-R $p$-forms of different $p$, via the Wess-Zumino
terms of ref. (\cite{wz}).  Magnetized branes, thus,
contribute simultaneously to tadpole cancellation conditions of
branes of different dimensionalities. Typically, a ``fat''
magnetic brane can replace a stack of localized lower-dimensional
branes, reducing the rank of the Chan-Paton group and
simultaneously yielding matter-field families determined by the
Landau-level degeneracies\cite{mag6,ted}. In the presence of at
least two kinds of (intersecting) $D$-branes and of compatible
shifts, it is possible to obtain four-dimensional {\it chiral}
models.  As an example, we shall discuss the magnetized
orientifold of the $w_1 w_2 p_3$~model in table~\ref{oldmodels},
but an exhaustive description of all the magnetized orientifolds
of the models in table~\ref{oldmodels} can be found in ref.
(\cite{magn}). The unoriented closed spectra of the $w_1 w_2
p_3$~model contain $N=1$ supergravity coupled to $7$ chiral
multiplets. The tadpole cancellation conditions, 
\bea
n + m  + \bar{m}  =  8 \ 2^{-{r \over 2}} \quad , \nonumber \\
d_1 + 2^{-{r_2 \over 2} - {r_3 \over 2}} \ | k_2 k_3 |
\ (m + \bar{m})   =  8 \  2^{-{r \over 2}} \quad , \label{tadpo}  \\
d_2  =  8 \ 2^{-{r \over 2}} \qquad ; \qquad  m = \bar{m} \quad, \nonumber 
\eea 
show how the open sector contains both uncharged
and magnetized $D9$-branes, whose numbers are labelled by $n$ and
$m$, together with two sets of $D5$ branes counted by $d_i$,
$i=1,2$ . The $k_i$ in eq. (\ref{tadpo}) are the integer Landau
level degeneracies of the internal magnetic field that lies along
the second and third tori, while $r=r_1+r_2+r_3$ is the total rank
of $B_{ab}$, whose three two-by-two sub-blocks have ranks $r_j$.
The gauge groups can be chosen to be orthogonal or symplectic for
the uncharged branes, while on the magnetized $D9$-branes the
gauge group is unitary.  The chiral open spectra are reported in
table~\ref{opunw1w2p3}, where $C$ denotes chiral multiplets,
$\eta_1$ is a free sign while $A$, $F$, $S$ and $Adj$ denote the
Antisymmetric, Symmetric, Fundamental and Adjoint representation
of the gauge group, respectively.
\begin{table}[t]
\caption{Open spectra of the $w_1w_2 p_3$-orientifold models.}
\begin{center}
\footnotesize
\begin{tabular}{|c|c|c|}\hline
{\rm Mult.}&{Number}&{\rm Reps. $(n,d_1,d_2,m)$}\\\hline
{\rm $C$}&{$3$}&{$(Adj,1,1,1),(1,Adj,1,1)$}\\
{}&{}&{$(1,1,Adj,1)$, $(1,1,1,Adj)$}\\\hline
{\rm $C$}&{$2^{r_2+r_3} \ 2 \ |k_2 \, k_3|$}&{$(F,1,1,F+\bar{F})$}\\
\hline
{\rm $C_L$}&{$2^{r_2+r_3} \ 4 \ |k_2 \, k_3| + 2^{\frac{r_2+r_3}{2}}
\eta_1 2 |k_2 \, k_3| + 2^{\frac{r_2}{2}} 2 |k_2|$}
&{$(1,1,1,A)$}\\\hline
{\rm $C_L$}&{$2^{r_2+r_3} \ 4 \ |k_2 \, k_3| - 2^{\frac{r_2+r_3}{2}}
\eta_1 2 |k_2 \, k_3| - 2^{\frac{r_2}{2}} 2 |k_2|$}
&{$(1,1,1,S)$}\\\hline
{\rm $C_L$}&{$2^{r_2+r_3} \ 4 \ |k_2 \, k_3| + 2^{\frac{r_2+r_3}{2}}
\eta_1 2 |k_2 \, k_3| - 2^{\frac{r_2}{2}} 2 |k_2|$}
&{$(1,1,1,\bar{A})$}\\\hline
{\rm $C_L$}&{$2^{r_2+r_3} \ 4 \ |k_2 \, k_3| - 2^{\frac{r_2+r_3}{2}}
\eta_1 2 |k_2 \, k_3| + 2^{\frac{r_2}{2}} 2 |k_2|$}
&{$(1,1,1,\bar{S})$}\\\hline
{\rm $C_L$}&{$2^{\frac{r_1+r_3}{2}+r_2} \ 4 \ |k_2|$}
&{$(1,1,F,F)$}\\\hline
\end{tabular}
\label{opunw1w2p3}
\end{center}
\end{table}
Adding open-string Wilson lines and brane-antibrane pairs, it is
possible to obtain models whose spectra are close to (some
extensions of) the Standard Model, and examples of this sort are
reviewed in other contributions to this volume. This is certainly
a step forward in the understanding of the possible
phenomenological implications of a brane-world-like scenario, even
if the study of the dynamical stability of this kind of vacua and
the lack of a ``vacuum selection principle'' are still serious
open problems.
\section*{Acknowledgments}

It is a pleasure to thank the Organizers of the SP2003 Conference for
the invitation and Marianna Larosa for the enjoyable collaboration.
I would also like to thank C. Angelantonj, M. Bianchi, E. Dudas,
Ya.S. Stanev and especially A. Sagnotti for
several interesting discussions.  This work was supported in
part by I.N.F.N., by the
E.C. RTN programs HPRN-CT-2000-00122 and HPRN-CT-2000-00148, by the
INTAS contract 99-1-590, by the MIUR-COFIN contract 2001-025492 and
by the NATO contract PST.CLG.978785.

\end{document}